\documentclass[12pt]{article}
\usepackage{amsmath}
\usepackage{bm}
\usepackage{color}
\usepackage{braket}
\usepackage{caption}
\usepackage{subcaption}
\usepackage{graphicx}

\begin{document}
\begin{center}
\begin{large}
{\bf Geometric measure of entanglement of multi-qubit graph states and its detection on a quantum computer}
\end{large}
\end{center}

\centerline {Kh. P. Gnatenko \footnote{E-Mail address: khrystyna.gnatenko@gmail.com}, N. A. Susulovska}
\medskip
\centerline {\small \it Ivan Franko National University of Lviv,}
\centerline {\small \it Professor Ivan Vakarchuk Department for Theoretical Physics,}
\centerline {\small \it 12 Drahomanov St., Lviv, 79005, Ukraine}

\abstract{Multi-qubit graph states generated by the action of controlled phase shift operators on a separable quantum state of a system, in which all the qubits are in arbitrary identical states, are examined. The geometric measure of entanglement of a qubit with other qubits is found for the graph states represented by arbitrary graphs. The entanglement depends on the degree of the vertex representing the qubit, the absolute values of the parameter of the phase shift gate and the parameter of state the gate is acting on. Also the geometric measure of entanglement of the graph states is quantified on the quantum computer $\textrm{ibmq\_athens}$. The results obtained on the quantum device are in good agreement with analytical ones.

}

\section{Introduction}

Studies of entanglement of quantum states and  its quantifying on a quantum computer have received much attention (see, for instance, \cite{Horodecki,Shimony,Behera,Scott,Horodecki1,Torrico,Sheng,Samar,Kuzmak,Kuzmak1,Gnatenko,Kuzmak2,Wang,Mooney} and references therein).   Entanglement  corresponds to  non-classical correlations between the subsystems and presupposes  that the state of a system cannot be factorized  \cite{Horodecki}. This physical phenomenon  plays a critical role in quantum information
in particular, in quantum cryptography, quantum teleportation (see, for example,  \cite{Horodecki,Feynman,Bennett,Bouwmeester,Ekert,Raussendorf,Lloyd,Buluta,Shi,Llewellyn,Huang,Yin,Jennewein,Karlsson}).

The geometric measure of entanglement proposed by Shimony \cite{Shimony} is defined as
a minimal squared Fubiny-Study distance $d_{FS}^2 (\vert\psi\rangle, \vert\psi_s\rangle)=1-|\langle\psi|\psi_s\rangle|^2$ between an entangled state   $\vert\psi\rangle$ and a set of separable pure states $\vert\psi_s\rangle$. It reads
\begin{eqnarray}
E(\vert\psi\rangle)=\min_{\vert\psi_s\rangle}(1-|\langle\psi|\psi_s\rangle|^2).
\end{eqnarray}

The authors of paper  \cite{Samar}  showed that  the geometric measure of entanglement of a spin one-half (or qubit) with a quantum system
 in a pure state $\vert\psi\rangle$ is entirely determined by the mean spin in this state.
  Namely the following relation is satisfied
 \begin{eqnarray}
E(\vert\psi\rangle)=\frac{1}{2}\left(1-|\langle{\bm \sigma}\rangle|\right),\label{ent}
\end{eqnarray}
$|\langle{\bm \sigma}\rangle|=\sqrt{\langle\sigma^x\rangle^2+\langle\sigma^y\rangle^2+\langle\sigma^z\rangle^2}$, here
$\sigma^x$, $\sigma^y$, $\sigma^z$  are the Pauli matrices, $\langle...\rangle=\bra \psi ...\ket \psi$.
Therefore, in order to quantify the geometric measure of entanglement the mean values  of the Pauli matrices have to be calculated. Quantum protocol for detecting $\langle\sigma^x\rangle$, $\langle\sigma^y\rangle$, $\langle\sigma^z\rangle$ on a quantum computer is presented in \cite{Kuzmak}.
 Mean value   $\langle\sigma^x \rangle$  can be represented as
$\langle\sigma^x \rangle=\langle \psi \vert \sigma^x\vert\psi\rangle=\langle \tilde\psi^y \vert \sigma^z\vert \tilde\psi^y\rangle=\vert \langle \tilde\psi^y \vert 0 \rangle \vert^2-\vert \langle \tilde\psi^y \vert 1 \rangle \vert^2,$
 where $\vert\tilde\psi^y\rangle=\exp(i\pi\sigma^y/4)\vert\psi\rangle$ and $\vert\langle\tilde\psi^y\vert 0 \rangle\vert^2$, $\vert\langle\tilde\psi^y\vert 1 \rangle \vert^2$ are  probabilities that define the result of measurement in the standard basis.  Thus, to  quantify $\langle\sigma^x \rangle$   the $RY(\pi/2)$ gate has to be applied  to the state of a qubit  before conducting the measurement in the standard basis (the state of the qubit has to be rotated around the $y$ axis by $\pi/2$).
 Similarly, to detect $\langle\sigma^y \rangle$ one has to apply the  $RX(\pi/2)$ gate and then measure the qubit in the standard basis
$  \langle\sigma^y \rangle=\langle \psi \vert \sigma^y\vert\psi\rangle=\langle \tilde\psi^x \vert \sigma^z\vert \tilde\psi^x\rangle=\vert \langle \tilde\psi^x \vert 0 \rangle \vert^2-\vert \langle \tilde\psi^x \vert 1 \rangle \vert^2$,
  here $\vert\tilde\psi^x\rangle=\exp(-i\pi\sigma^x/4)\vert\psi\rangle$ .
 Lastly, for $\sigma^z$  we have
$\langle\sigma^z \rangle=\langle \psi \vert \sigma^z\vert\psi\rangle=\vert \langle\psi\vert 0 \rangle \vert^2-\vert \langle \psi \vert 1 \rangle\vert^2.$
 This way one can obtain the values  $\langle\sigma^x\rangle$, $\langle\sigma^y\rangle$, $\langle\sigma^z\rangle$ on the basis of the results of measurement of the qubit in the standard basis \cite{Kuzmak}.

Graph states are quantum states that can be represented by graphs \cite{Wang,Mooney,Schlingemann,Bell,Mazurek,Markham,Qian,Shettell,Hein,Guhne}. These states have various applications in quantum information, for instance in
quantum error correction \cite{Schlingemann,Bell,Mazurek},
quantum cryptography \cite{Markham,Qian} and practical quantum metrology in the presence of noise \cite{Shettell}.
Much attention has been devoted to examining multi-qubit graph states generated by 2-qubit controlled-Z operators acting on a separable quantum state of the system, in which all  qubits are in  the state $\ket{+}=(\ket{0}+\ket{1})/\sqrt{2}$ (see, for example, \cite{Wang,Mooney,Cabello,Alba,Mezher,Akhound,Haddadi} and references therein). The state represented by undirected graph $G(V,E)$ ($V$, $E$ denote vertices and edges of the graph,respectively) reads
 \begin{eqnarray}
		\ket{\psi_G} = \prod_{(i,j) \in E} CZ_{ij}(\phi) \ket{+}^{\otimes V}. \label{eq:2:1z}
	\end{eqnarray}
Here $CZ_{ij}(\phi)$ is the controlled-Z gate acting on the states of qubits $q[i]$, $q[j]$.

The authors of paper \cite{Wang} studied graph states (\ref{eq:2:1z}) corresponding to rings on IBM's quantum computer ibmqx5 and showed that the 16-qubit  quantum computer can be fully entangled. In \cite{Mooney}  the entanglement of  graph states was examined on the basis of calculations on the quantum computer IBM Q Poughkeepsie. Entanglement of graph states  of spin systems
generated by the operator of evolution  with Ising Hamiltonian was examined analytically and  on the 5-qubit quantum computer IBM Q Valencia in \cite{Gnatenko}. It was shown that the entanglement of a spin with other spins in the graph state is related to the degree of the vertex representing the spin \cite{Gnatenko}.

 In the present paper we study graph states obtained  as a result of action of controlled phase shift operators on a separable quantum state of the system, in which all the qubits are in arbitrary identical states
 \begin{eqnarray}
		\ket{\psi_G(\phi,\alpha,\theta)} = \prod_{(i,j) \in E} CP_{ij}(\phi) \ket{\psi(\alpha,\theta)}^{\otimes V}, \label{eq:2:1}\\
		\ket{\psi(\alpha,\theta)} = \cos \frac{\theta}{2} \ket{0}+ e^{i\alpha} \sin \frac{\theta}{2} \ket{1}, \label{eq:2:2}
	\end{eqnarray}
here $CP_{ij}(\phi)$ is the controlled phase shift gate that acts on the qubits $q[i]$, $q[j]$.
 State (\ref{eq:2:2}) is an arbitrary one-qubit state, $\theta\in[0,\pi]$, $\phi\in[0,2\pi]$.
In a particular case $\phi=\pi$, $\alpha=0$, $\theta = \pi/2$ state (\ref{eq:2:1}) coincides with (\ref{eq:2:1z}).
  We find the expression for the entanglement of a qubit with other qubits in graph state (\ref{eq:2:1}) represented by an arbitrary graph. It is  shown that the entanglement is determined by the absolute values of the parameters $\phi$ and $\theta$ as well as the degree of the vertex representing the qubit in the graph.
The entanglement of graph states is also studied on  IBM's quantum computer $\textrm{ibmq\_athens}$.

The paper is organized as follows. In Section 2 an expression for the geometric measure of entanglement of a qubit with other qubits in graph state (\ref{eq:2:1}) is found. In Section 3 we present results of quantum computations on $\textrm{ibmq\_athens}$ for the entanglement of the graph states (\ref{eq:2:1}) represented by the chain, the claw and the complete graphs.  Conclusions are made in Section 4.

\section{Geometric measure of entanglement of multi-qubit graph states}

According to the result (\ref{ent}) in order to find the geometric measure of entanglement of the qubit $q[l]$ with other qubits in graph state (\ref{eq:2:1}) one has to calculate $\langle{\bm \sigma}_l\rangle=\langle \psi_G (\phi,\alpha,\theta)\vert{\bm \sigma}_l\vert\psi_G (\phi,\alpha,\theta)\rangle$.
Note that accurate to the phase factor the quantum state $\ket{\psi(\alpha,\theta)}$ can be prepared by the action of the rotation operators $RZ(\alpha)$, $RY(\theta)$ on  $\ket 0$. We have
\begin{eqnarray}
\ket{\psi(\alpha,\theta)} =e^{i\frac{\alpha}{2}}RZ(\alpha)RY(\theta)\ket 0=\nonumber\\=e^{i\frac{\alpha}{2}}e^{-i\frac{\alpha}{2}\sigma^z}e^{-i\frac{\theta}{2}\sigma^y}\ket 0.\label{eq:2:3}
\end{eqnarray}

The controlled phase shift gate can be represented as
\begin{eqnarray}
CP_{ij}(\phi)=|0\rangle_i{}_i\langle0| \hat{1}_j+|0\rangle_i{}_i\langle0| P_j(\phi)
=\nonumber\\=e^{\frac{i \phi}{4} (\hat{1}_i- \sigma^z_i) (\hat{1}_j - \sigma^z_j)},
\end{eqnarray}
 where $\hat{1}_i$ is the unit operator,  $P_j(\phi)$ is the phase gate acting on the state of the qubit $q[j]$, $P_j(\phi)=|0\rangle_j{}_j\langle0|+e^{i\phi}|0\rangle_j{}_j\langle0|$.

Thus, for $\braket{\sigma^x_l}$ we obtain
	\begin{eqnarray}
\braket{\sigma^x_l}  = \bra{\psi_0} \prod_{q\in V}  e^{i\frac{\theta}{2} \sigma^y_q} e^{i\frac{ \alpha}{2} \sigma^z_q} \prod_{(j,k) \in E} (CP_{jk}(\phi))^+ \times\nonumber\\ \times\sigma^x_l \prod_{(m,n) \in E} CP_{mn}(\phi) \prod_{p \in V} e^{-i\frac{ \alpha}{2} \sigma^z_p} e^{-i\frac{ \theta}{2} \sigma^y_p} \ket{\psi_0}, \label{eq:2:5}
	\end{eqnarray}
here we use notation ${\psi_0}=\ket{0}^{\otimes V}$.
Taking into account that operators $\sigma^x_l$, $\sigma^z_l$ anticommute ($\{\sigma^z_l, \sigma^x_l\}=0$), we can write
\begin{eqnarray}
 \prod_{(j,k) \in E} (CP_{jk}(\phi))^+ \sigma^x_l \prod_{(m,n) \in E} CP_{mn}(\phi) =\nonumber\\=\prod_{(j,k) \in E} e^{-i\frac{ \phi}{4} (\hat{1}_j- \sigma^z_j) (\hat{1}_k - \sigma^z_k)}\sigma^x_l \times \nonumber\\ \times\prod_{(m,n) \in E} e^{i\frac{ \phi}{4} (\hat{1}_m-\sigma^z_m) (\hat{1}_n - \sigma^z_n)}=\nonumber\\=e^{i\frac{\phi}{2} n_l \sigma^z_l} e^{-i \frac{\phi}{2} \sum_{j \in N_G(l)} \sigma^z_j \sigma^z_l}  \sigma^x_l,
\end{eqnarray}
here $n_l$ is a degree of the vertex representing the qubit $q[l]$ in the graph (a number of edges incident to the vertex),  $N_G(l)$ is a neighborhood of the vertex $l$ (a set of vertices adjacent to the vertex $l$).
Hence, for $\braket{\sigma^x_l}$ we obtain
	\begin{eqnarray}
\braket{\sigma^x_l} =\nonumber\\ = \bra{\psi_0} \prod_{q \in N_G[l]}  e^{i\frac{\theta}{2} \sigma^y_q} e^{i\frac{ \alpha}{2} \sigma^z_q}e^{i\frac{\phi}{2} n_l \sigma^z_l} e^{-i \frac{\phi}{2} \sum_{j \in N_G(l)}\sigma^z_j \sigma^z_l}   \times \nonumber\\ \times \sigma^x_l
\prod_{p\in N_G[l]} e^{-i\frac{ \alpha}{2} \sigma^z_p} e^{-i\frac{ \theta}{2} \sigma^y_p} \ket{\psi_0}=
 \sin \theta \; \textrm{Re}\;z,  \label{eq:2:6}
	\end{eqnarray}
where $N_G[l]$ is a closed neighbourhood of the vertex $l$ (a set of vertices adjacent to the vertex  $l$ and the vertex $l$). We use notation $z$ for complex number
\begin{eqnarray}
z=e^{-i(\alpha + \frac{\phi}{2} n_l)} \left(\cos \frac{\phi}{2} + i \sin \frac{\phi}{2} \cos \theta \right)^{n_l}.\label{zz}
\end{eqnarray}

Similarly, for  $\braket{\sigma^y_l}$ we find
	\begin{eqnarray}
\braket{\sigma^y_l} = \bra{\psi_G} \sigma^y_l \ket{\psi_G} = \nonumber\\
\bra{\psi_0} \prod_{q\in N_G[l]}  e^{i\frac{\theta}{2} \sigma^y_q} e^{i\frac{ \alpha}{2} \sigma^z_q}e^{i\frac{\phi}{2} n_l \sigma^z_l} e^{-i \frac{\phi}{2} \sum_{j \in N_G(l)} \sigma^z_j \sigma^z_l}  \times \nonumber\\ \times \sigma^y_l
\prod_{p\in N_G[l]} e^{-i\frac{ \alpha}{2} \sigma^z_p} e^{-i\frac{ \theta}{2} \sigma^y_p} \ket{\psi_0}
= - \sin \theta \; \textrm{Im}\;z, \label{eq:2:20}
	\end{eqnarray}
number $z$ is given by (\ref{zz}).
Mean value  $\braket{\sigma^z_l}$ reads
	\begin{eqnarray}
		\braket{\sigma^z_l} &= \bra{\psi_G} \sigma^z_l \ket{\psi_G} = \bra{\psi_0} e^{i \frac{\theta}{2} \sigma^y_l} \; \sigma^z_l  e^{-i \frac{\theta}{2} \sigma^y_l} \ket{\psi_0}=\nonumber\\=\cos \theta. \label{eq:2:22}
	\end{eqnarray}
	
Finally, on the basis of (\ref{ent}) for the geometric measure of entanglement of the qubit $q[l]$ with other qubits in graph state (\ref{eq:2:1}) we obtain the following expression
	\begin{eqnarray}
		E_l = \frac{1}{2} \left(1 - \sqrt{\sin^2\theta \;|z|^2 + \cos^2\theta} \right)=
 \frac{1}{2}-\nonumber\\-\frac{1}{2}\sqrt{\sin^2\theta \left(\cos^2\frac{\phi}{2} + \sin^2\frac{\phi}{2} \cos^2 \theta \right)^{n_l} + \cos^2 \theta }.\label{eq:2:28}
	\end{eqnarray}

Note that the geometric measure of entanglement of the qubit $q[l]$ with other qubits in graph state (\ref{eq:2:1})   depends on the  degree of the vertex $n_l$ representing $q[l]$  in the graph, the absolute values of the parameter of the controlled phase gate  $\phi$ and the parameter of state (\ref{eq:2:2}) $\theta$. It does not depend on the value of $\alpha$.

\section{Preparation of multi-qubit graph states and detection of their entanglement on a quantum computer}

We quantify the geometric measure of entanglement of graph states (\ref{eq:2:1}) on IBM's  5-qubit quantum computer  $\textrm{ibmq\_athens}$ \cite{kk}.
The structure of the quantum computer is presented in Fig. 1, where the arrows link qubits to which the CNOT gate can
be directly applied.
	\begin{figure}[h!]
		\centering
	\includegraphics[scale=0.33]{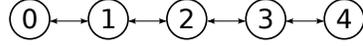}
		\caption{ Connectivity map of IBM's quantum computer $\textrm{ibmq\_athens}$. Arrows link qubits to which the CNOT gate can
be directly applied.}
		\label{fig:1}
	\end{figure}

\begin{figure}[h!]
\begin{center}
\subcaptionbox{\label{ff1}}{\includegraphics[scale=0.3, angle=0.0, clip]{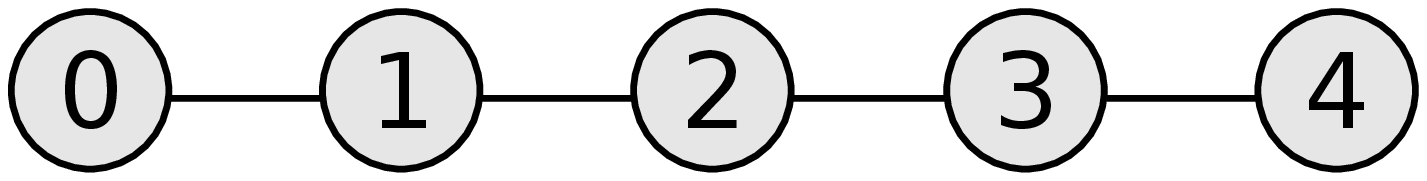}}
\hspace{1cm}
\subcaptionbox{\label{ff3}}{\includegraphics[scale=0.3, angle=0.0, clip]{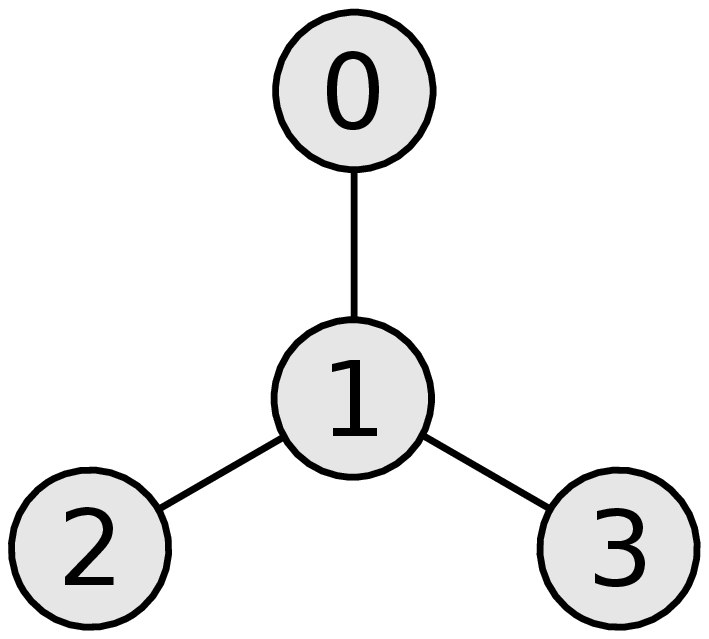}}
\hspace{1cm}
\subcaptionbox{\label{ff2}}{\includegraphics[scale=0.3, angle=0.0, clip]{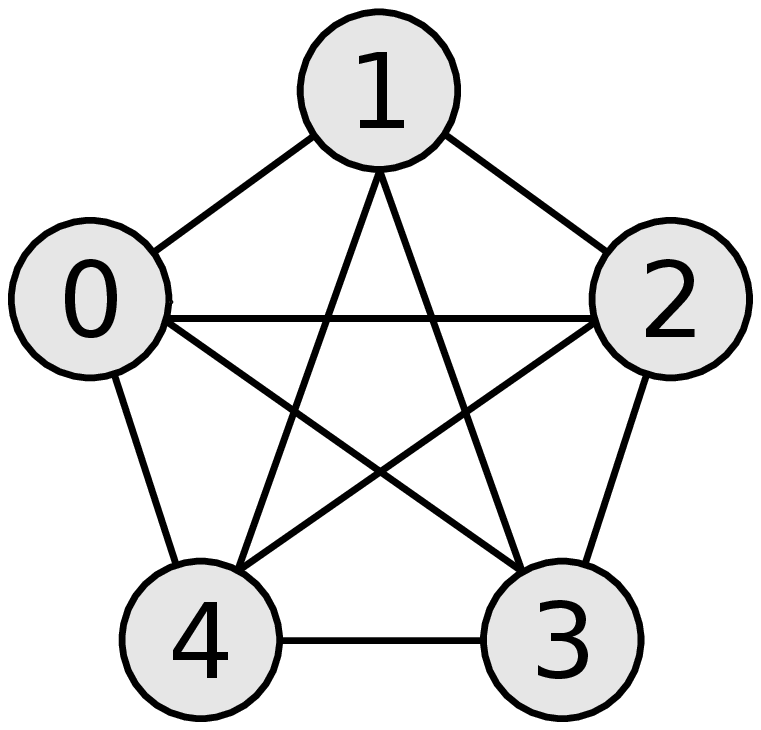}}
\caption{Graphs corresponding to graph states  $\ket{\psi_G^{(1)}(\phi,\alpha,\theta)}$ (\ref{eq:2:13})  (a), $\ket{\psi_G^{(2)}(\phi,\alpha,\theta)}$ (\ref{eq:2:14}) (b), $\ket{\psi_G^{(3)}(\phi,\alpha,\theta)}$ (\ref{eq:2:15}) (c).}
		\label{fig:2}
\end{center}
	\end{figure}

We consider graph state  (\ref{eq:2:1}) corresponding to the graph with the structure of  $\textrm{ibmq\_athens}$. It reads
\begin{eqnarray}
\ket{\psi_G^{(1)}(\phi,\alpha,\theta)} =\nonumber\\=CP_{01}(\phi)CP_{12}(\phi)CP_{23}(\phi)CP_{34}(\phi) \ket{\psi(\alpha,\theta)}^{\otimes 5}. \label{eq:2:13}
\end{eqnarray}
Degrees of vertices  in the graph (see Fig.\ref{fig:2} (a)) corresponding to state (\ref{eq:2:13})  are the following $\textrm{deg}(V_0)=\textrm{deg}(V_4)=1$, $\textrm{deg}(V_1)=\textrm{deg}(V_2)=\textrm{deg}(V_3)=2$, where $V_i$ is the vertex representing the qubit $q[i]$.
We also study graph states  (\ref{eq:2:1})  associated with the claw and the complete graphs (see Fig.\ref{fig:2} (b), (c)) and determine the geometric measure of entanglement of qubits represented by vertices with degrees 3 and 4.
These graph states are defined as follows
\begin{eqnarray}
\ket{\psi_G^{(2)}(\phi,\alpha,\theta)} =CP_{10}(\phi)CP_{12}(\phi)CP_{13}(\phi)\ket{\psi(\alpha,\theta)}^{\otimes 4}, \label{eq:2:14}\\
\ket{\psi_G^{(3)}(\phi,\alpha,\theta)} =\mathop{\prod^4_{i,j=0}}\limits_{i\neq j}CP_{ij}(\phi)\ket{\psi(\alpha,\theta)}^{\otimes 5}. \label{eq:2:15}
\end{eqnarray}
State (\ref{eq:2:14}) corresponds to the claw graph (see Fig.\ref{fig:2} (b)), which is  the most simple graph  with maximal vertex degree $\textrm{deg}(V_1)=3$. State (\ref{eq:2:15}) can be represented by the complete graph (see  Fig.\ref{fig:2} (c)). In this case  $\textrm{deg}(V_i)=4$, $i=(0,..,4)$.

To quantify dependence of the geometric measure of entanglement of a graph state on the angle $\phi$  we fix parameter $\theta$ as $\theta=\pi/2$.
Taking into account that expression  (\ref{eq:2:28}) obtained in the previous section does not depend on the value of $\alpha$, for convenience we set $\alpha=0$.
In this case from (\ref{eq:2:2}) we obtain $\ket{\psi(0,\pi/2)} =\ket+$ and the graph state reads
\begin{eqnarray}
		\ket{\psi_G(\phi,0,\pi/2)} = \prod_{(i,j) \in E} CP_{ij}(\phi) \ket{+}^{\otimes V}. \label{eq:2:11}
	\end{eqnarray}
Note that for $\phi=\pi$ state (\ref{eq:2:11}) coincides with  (\ref{eq:2:1z}).

In case of fixing parameter $\phi$ as $\phi=\pi$ for $\alpha=0$  graph state (\ref{eq:2:1}) transforms to
\begin{eqnarray}
\ket{\psi_G(\pi,0,\theta)} = \prod_{(i,j) \in E} CZ_{ij} \ket{\psi(0,\theta)}^{\otimes V}. \label{eq:2}
\end{eqnarray}

\begin{figure}[!!h]
\begin{center}
\subcaptionbox{\label{ff1}}{\includegraphics[scale=0.5, angle=0.0, clip]{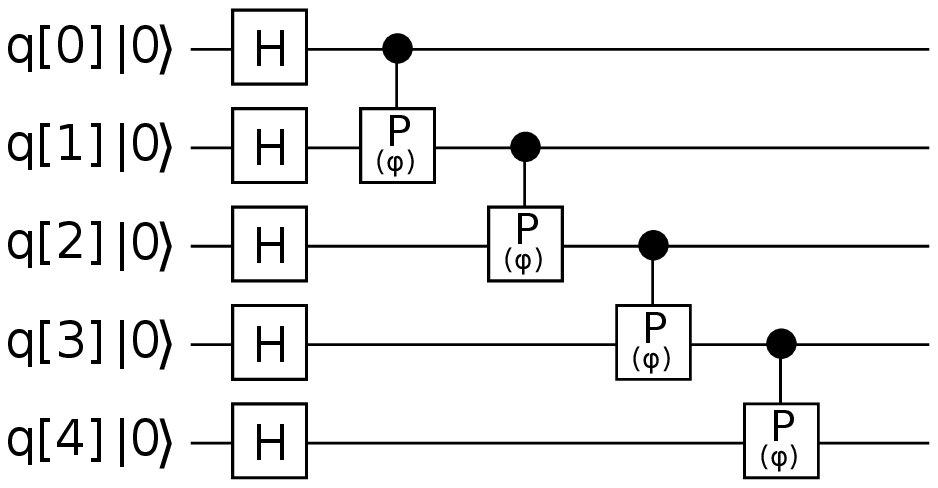}}
\hspace{1cm}
\subcaptionbox{\label{ff3}}{\includegraphics[scale=0.5, angle=0.0, clip]{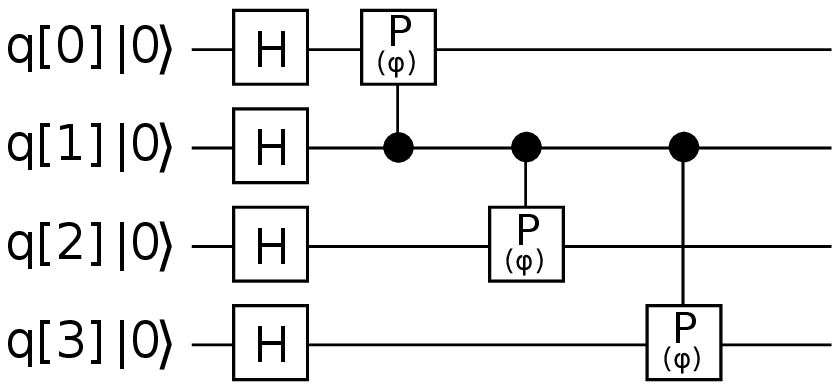}}
\subcaptionbox{\label{ff2}}{\includegraphics[scale=0.5, angle=0.0, clip]{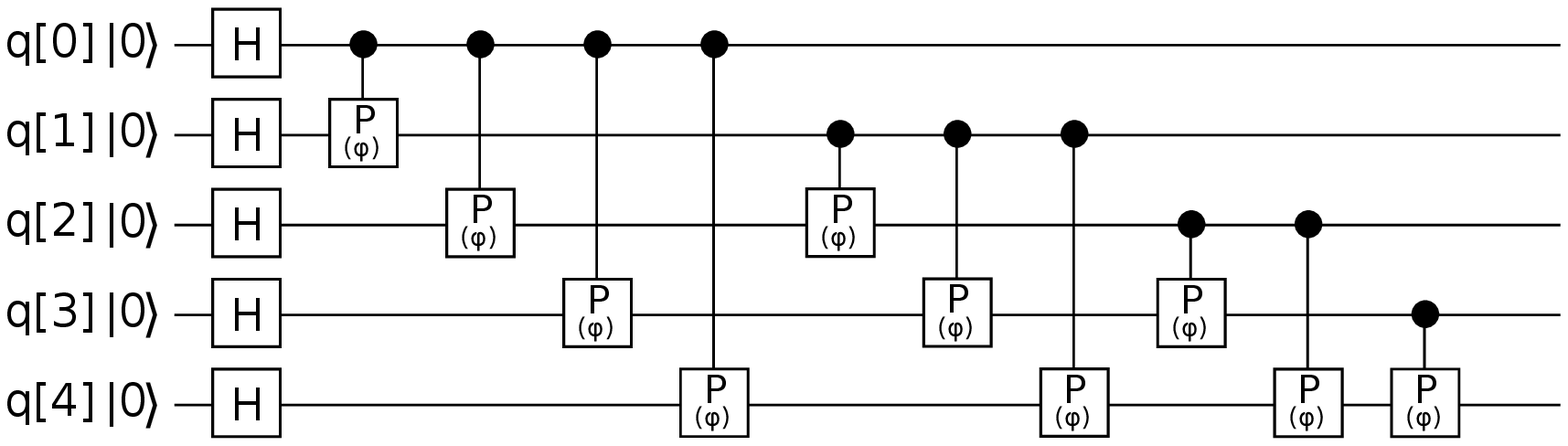}}
\end{center}
\caption{Quantum protocols for preparing graph states $\ket{\psi_G^{(1)}(\phi,0,\pi/2)}$ (\ref{eq:2:13})  (a), $\ket{\psi_G^{(2)}(\phi,0,\pi/2)}$ (\ref{eq:2:14})  (b),  $\ket{\psi_G^{(3)}(\phi,0,\pi/2)}$  (\ref{eq:2:15}) (c).}
		\label{fig:5}
\end{figure}

\begin{figure}[!!h]
\begin{center}
\subcaptionbox{\label{ff1}}{\includegraphics[scale=0.5, angle=0.0, clip]{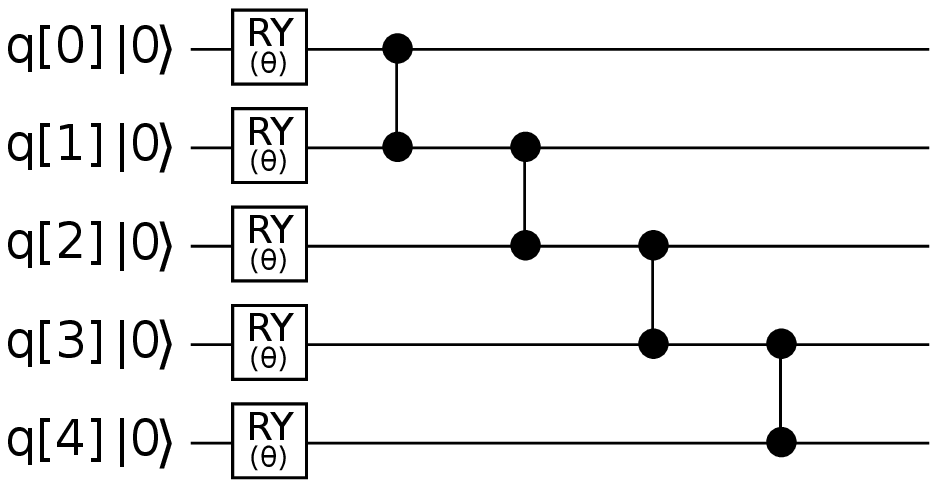}}
\hspace{1cm}
\subcaptionbox{\label{ff3}}{\includegraphics[scale=0.5, angle=0.0, clip]{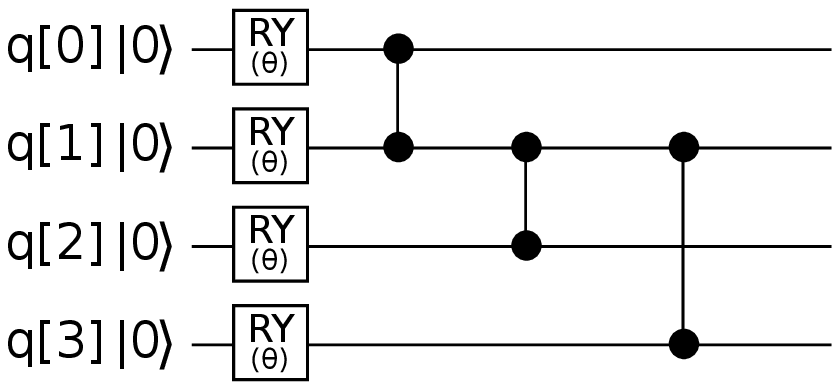}}
\subcaptionbox{\label{ff2}}{\includegraphics[scale=0.5, angle=0.0, clip]{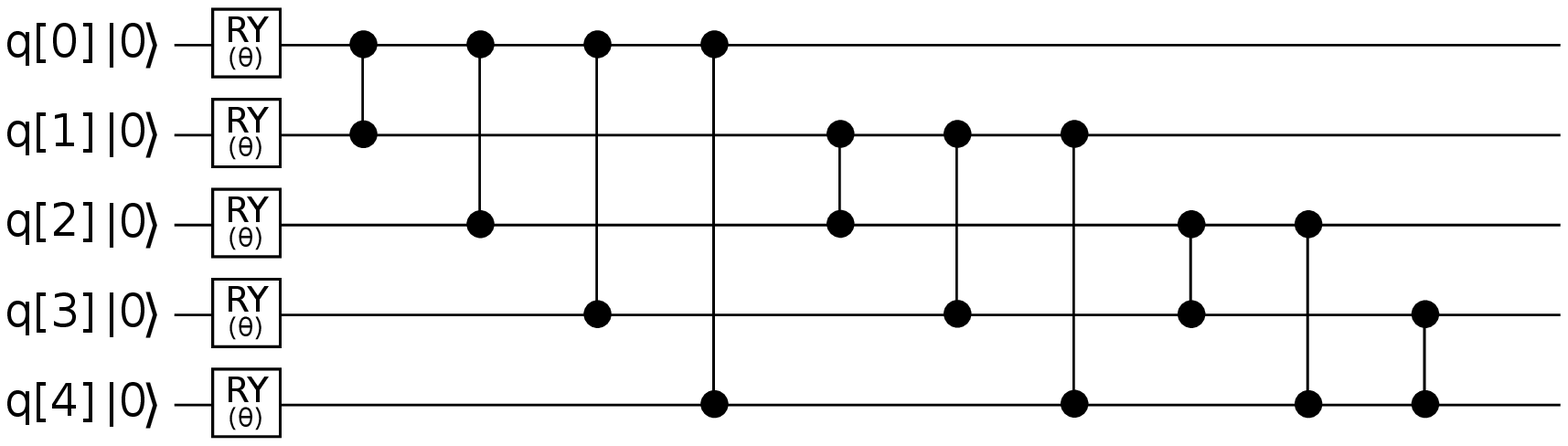}}
\end{center}
\caption{Quantum protocols for preparing graph states $\ket{\psi_G^{(1)}(\pi,0,\theta)}$ (\ref{eq:2:13})  (a), $\ket{\psi_G^{(2)}(\pi,0,\theta)}$ (\ref{eq:2:14})  (b),  $\ket{\psi_G^{(3)}(\pi,0,\theta)}$  (\ref{eq:2:15}) (c).}
		\label{fig:6}
\end{figure}

Quantum protocols for preparing graph states $\ket{\psi_G(\phi,0,\pi/2)}$, $\ket{\psi_G(\pi,0,\theta)}$ corresponding to the chain, the claw and the complete graphs (see (\ref{eq:2:13}), (\ref{eq:2:14}), (\ref{eq:2:15}), respectively) are presented in
Figs. \ref{fig:5}, \ref{fig:6}.

We detect entanglement of qubits corresponding to vertices with degrees 1, 2, 3, 4  in graph states $\ket{\psi_G^{(1)}(\phi,0,\theta)}$ (\ref{eq:2:13}), $\ket{\psi_G^{(2)}(\phi,0,\theta)}$ (\ref{eq:2:14}),  $\ket{\psi_G^{(3)}(\phi,0,\theta)}$  (\ref{eq:2:15}) on the quantum computer $\textrm{ibmq\_athens}$. For this purpose the graph states have been prepared with quantum protocols presented in Figs. \ref{fig:5}, \ref{fig:6} and the mean values $\braket{\sigma_l^x}$, $\braket{\sigma_l^y}$, $\braket{\sigma_l^z}$ have been measured using protocols presented in \cite{Kuzmak}.

\begin{figure}[h!]
 \begin{center}
\subcaptionbox{\label{ff1}}{\includegraphics[scale=0.35, angle=0.0, clip]{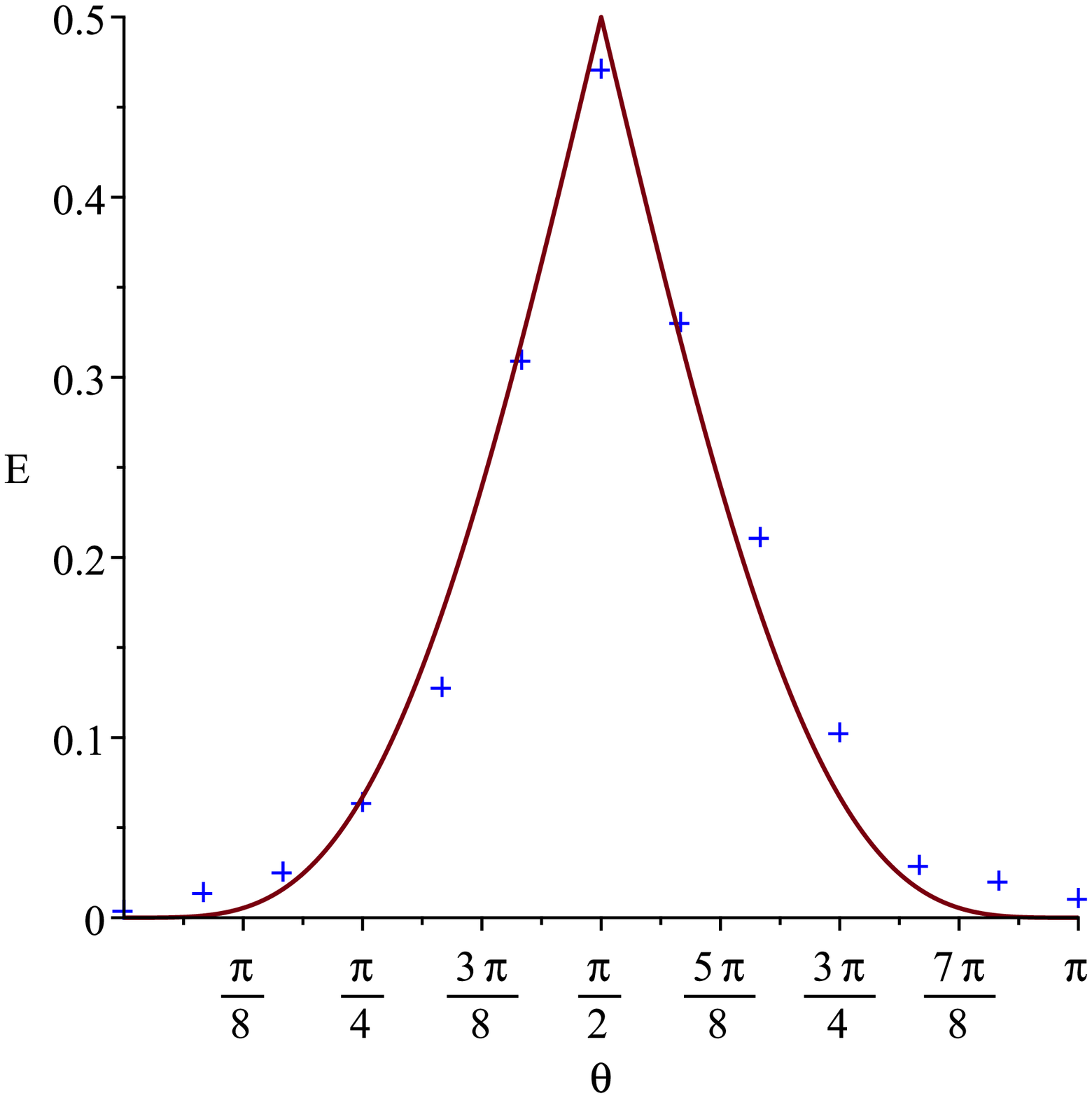}}
\subcaptionbox{\label{ff3}}{\includegraphics[scale=0.35, angle=0.0, clip]{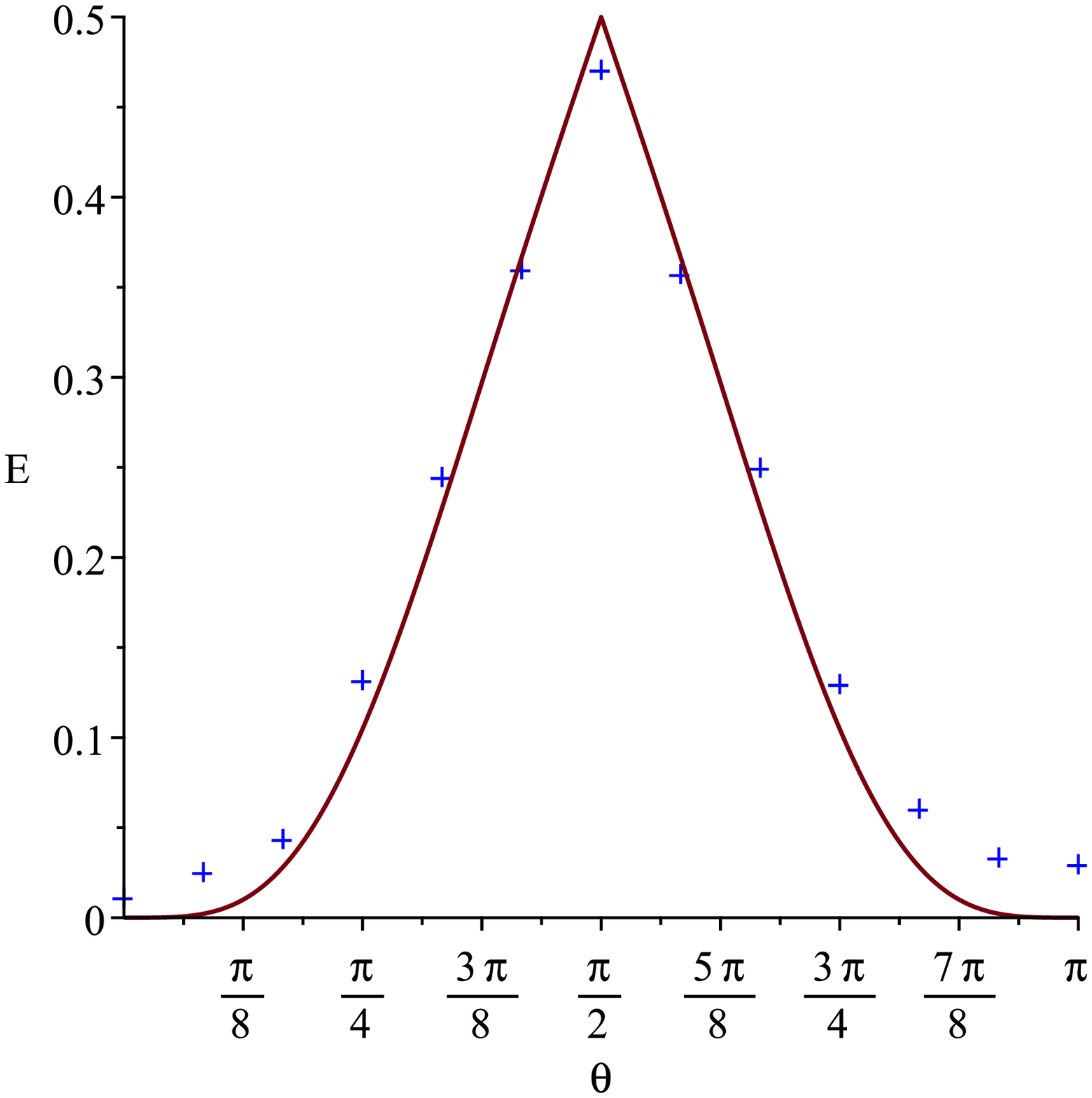}}
\subcaptionbox{\label{ff2}}{\includegraphics[scale=0.35, angle=0.0, clip]{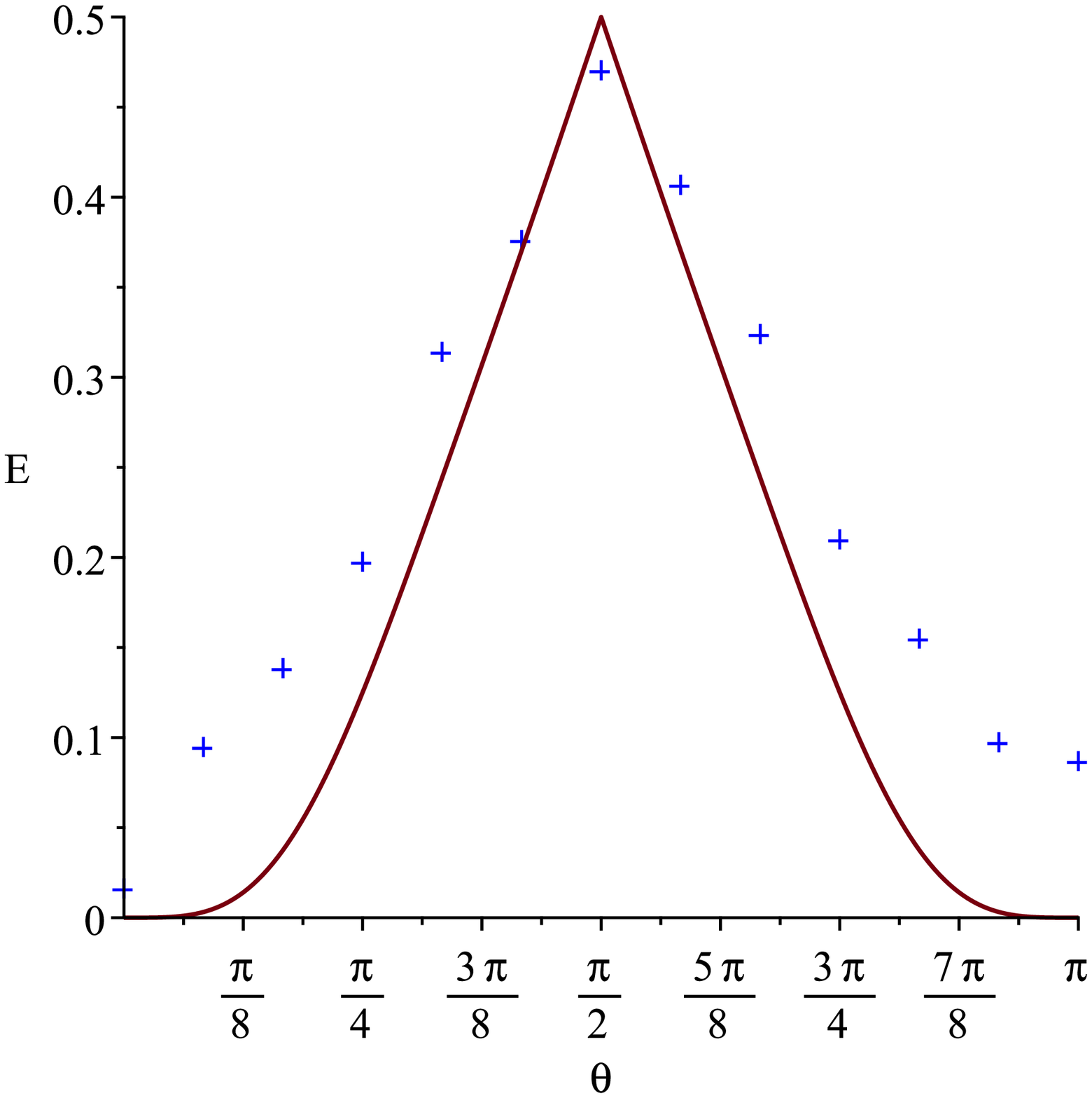}}
\subcaptionbox{\label{ff2}}{\includegraphics[scale=0.35, angle=0.0, clip]{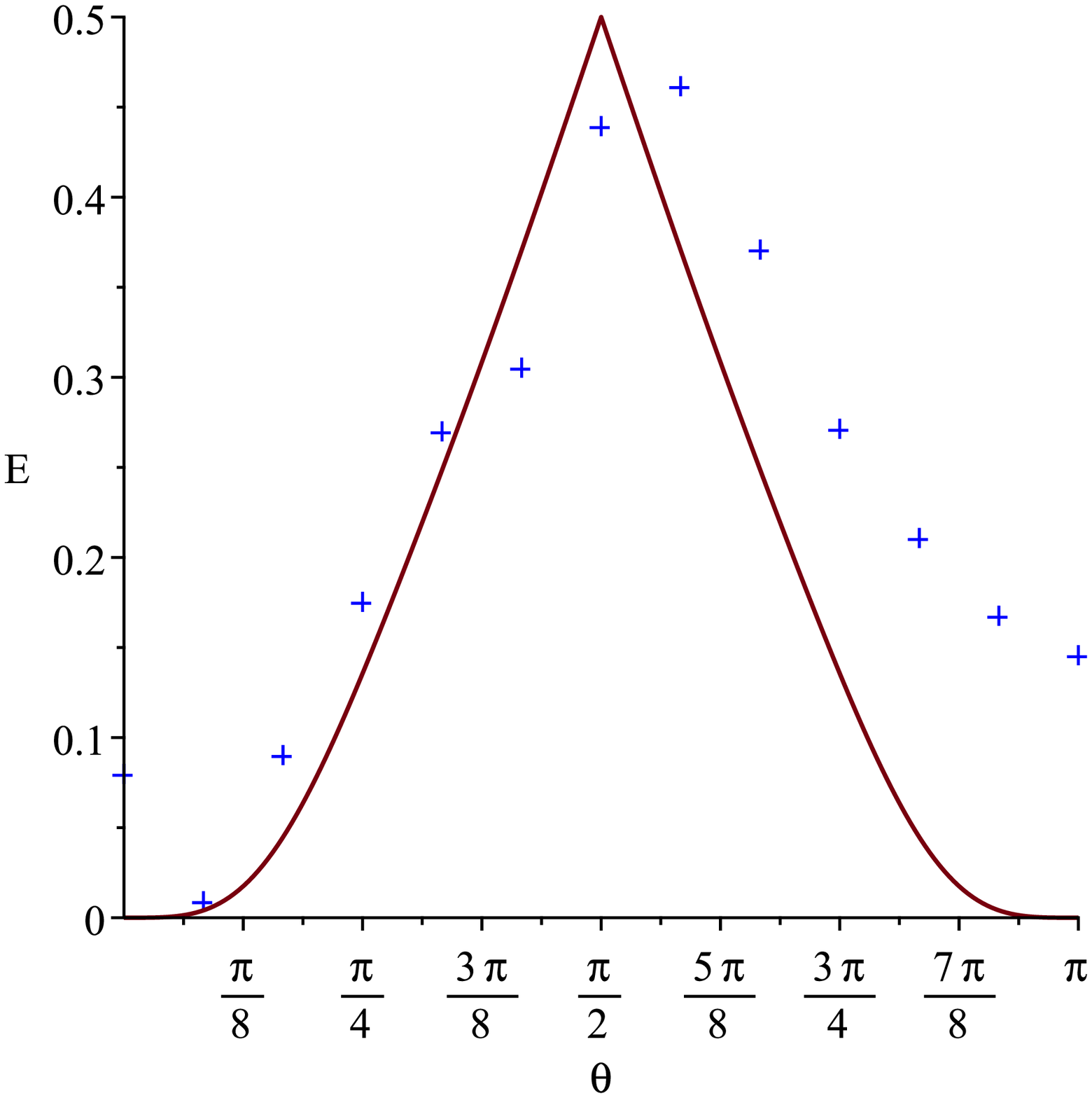}}
\end{center}
\caption{Results of detecting geometric measure of entanglement on  $\textrm{ibmq\_athens}$ quantum computer  (marked by crosses) and analytical results (line) for qubit q[0] (a) and qubit q[1]  (b) with other qubits  in  state $\ket{\psi_G^{(1)}(\pi,0,\theta)}$ (\ref{eq:2:13}) for different values of $\theta$, for qubit q[1]  with other qubits in  state $\ket{\psi_G^{(2)}(\pi,0,\theta)}$ (\ref{eq:2:14}) (c) and for qubit q[0]  with other qubits in  state   $\ket{\psi_G^{(3)}(\pi,0,\theta)}$ (\ref{eq:2:15}) (d) for different values of $\theta$.}
		\label{fig:7}
	\end{figure}

\begin{figure}[h!]
 \begin{center}
\subcaptionbox{\label{ff1}}{\includegraphics[scale=0.35, angle=0.0, clip]{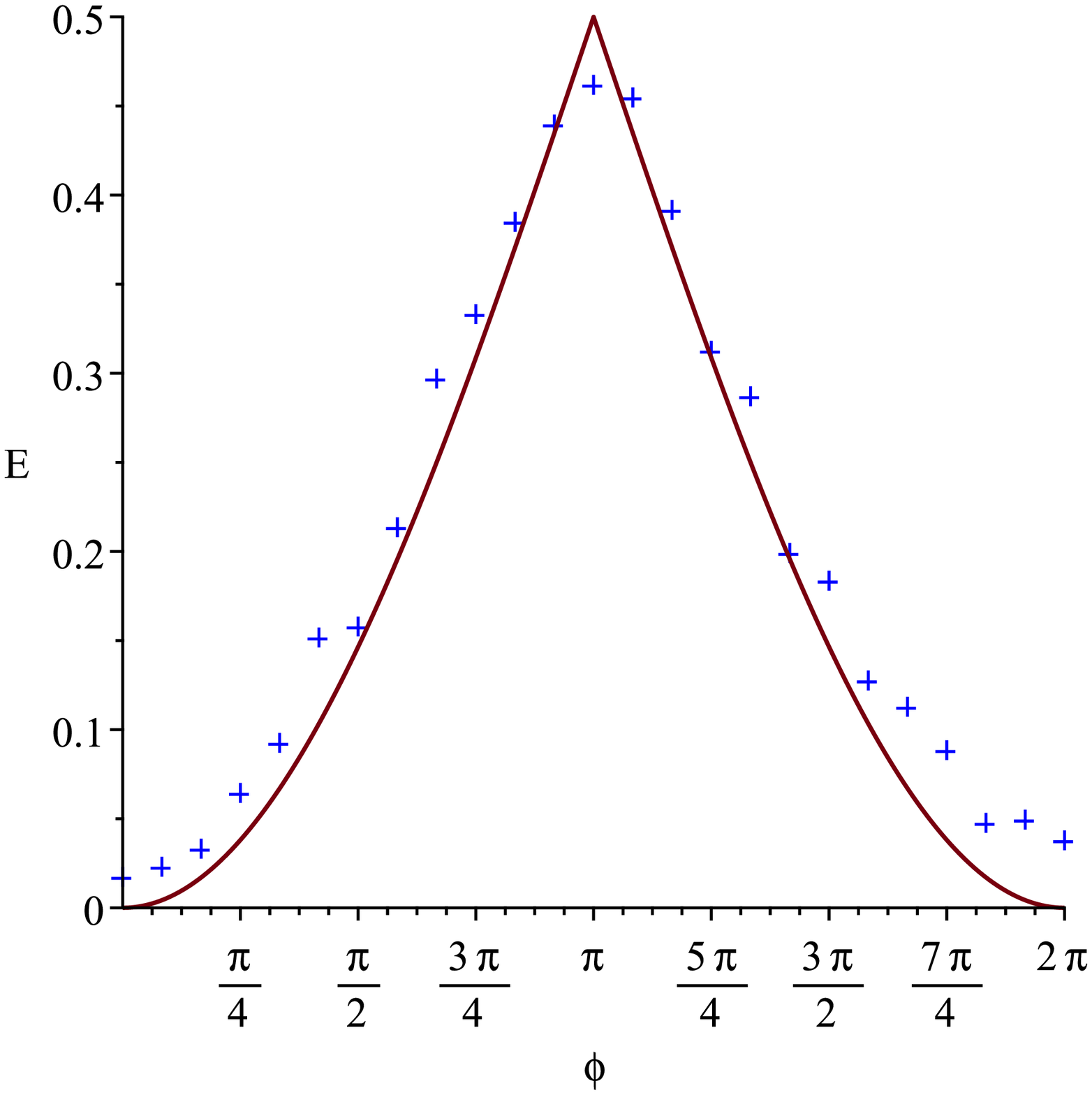}}
\subcaptionbox{\label{ff3}}{\includegraphics[scale=0.35, angle=0.0, clip]{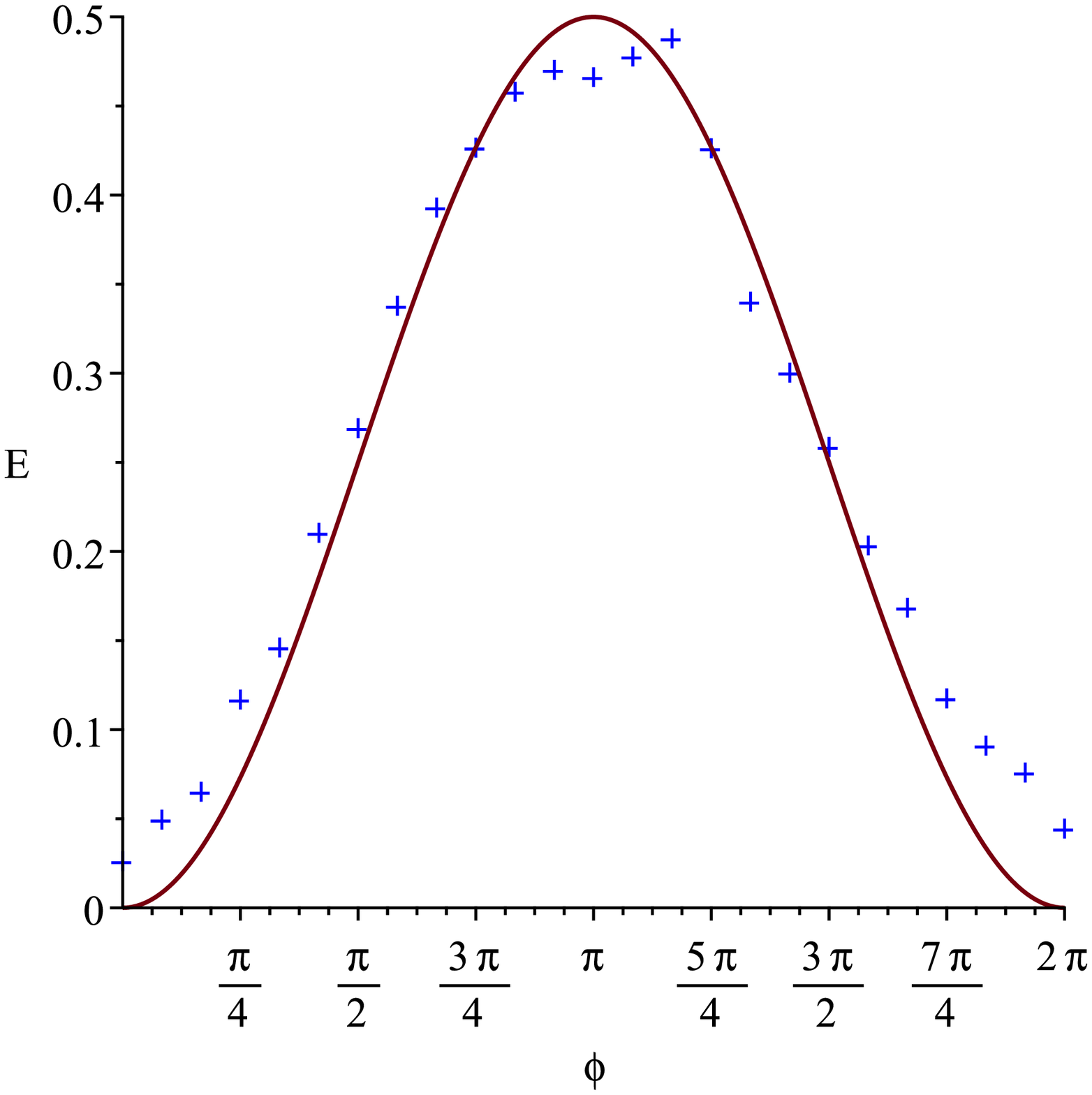}}
\subcaptionbox{\label{ff2}}{\includegraphics[scale=0.35, angle=0.0, clip]{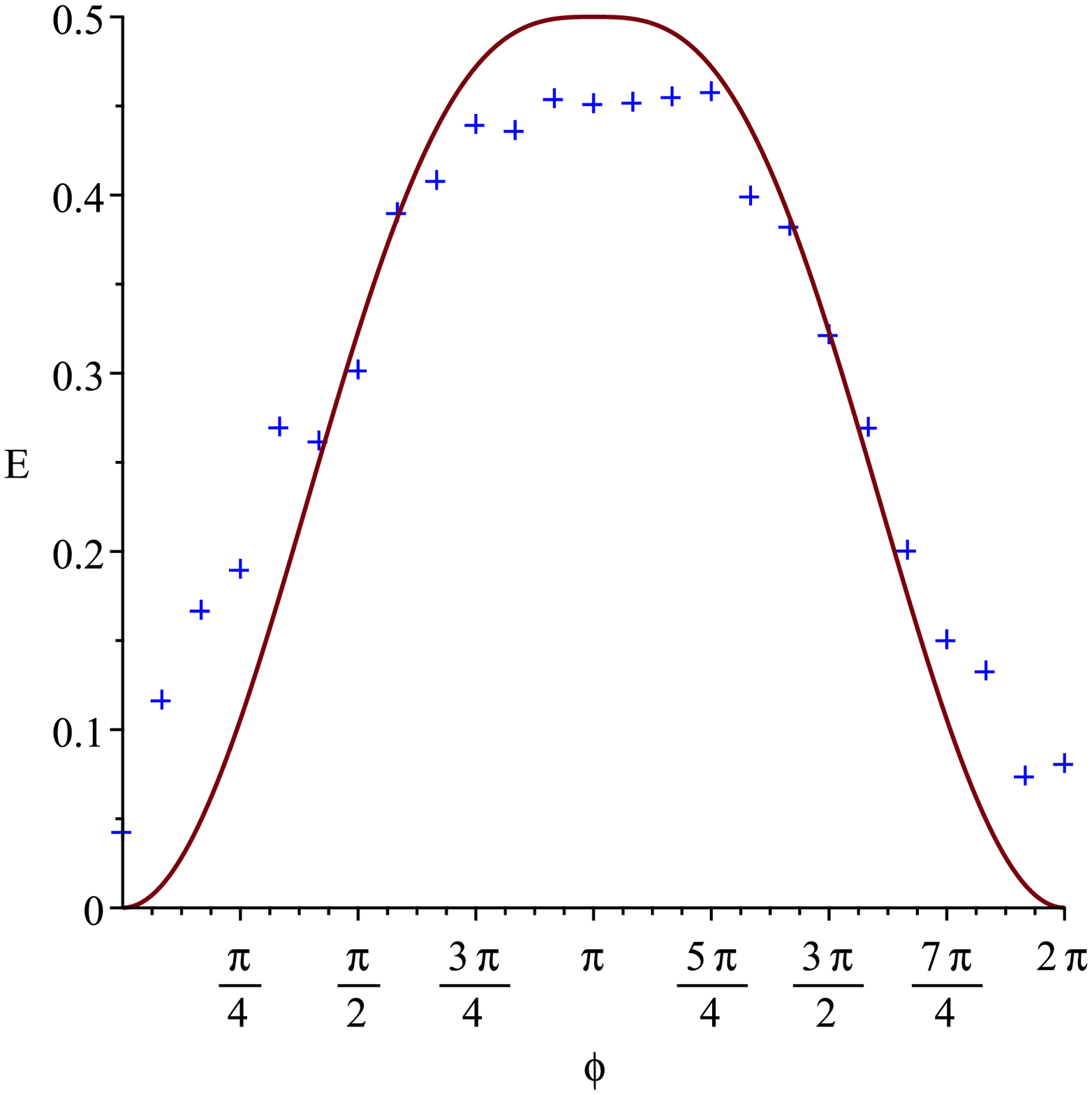}}
\subcaptionbox{\label{ff2}}{\includegraphics[scale=0.35, angle=0.0, clip]{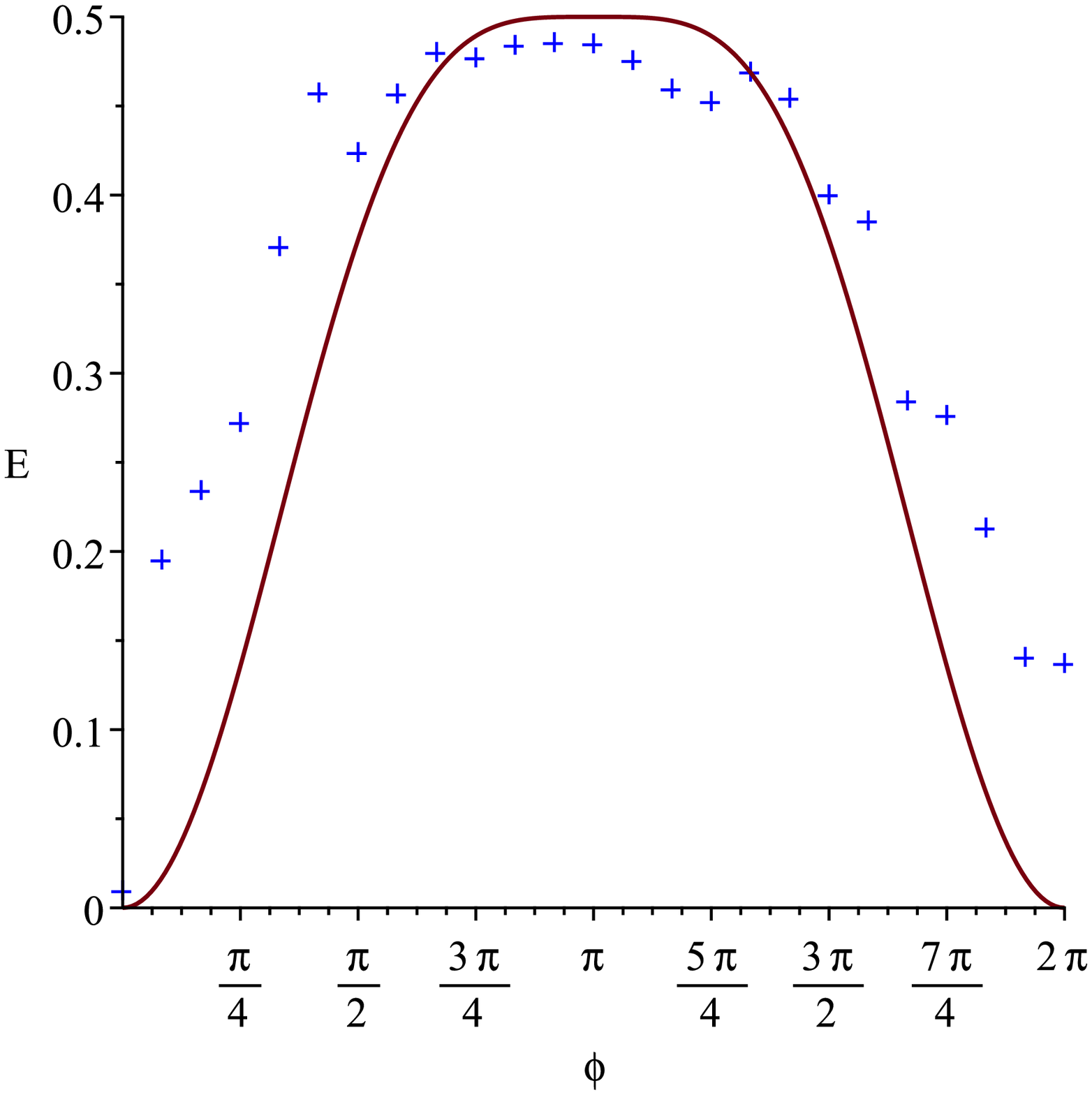}}
\end{center}
\caption{Results of detecting geometric measure of entanglement on  $\textrm{ibmq\_athens}$ quantum computer  (marked by crosses) and analytical results (line) for qubit q[0] (a) and qubit q[1]  (b) with other qubits  in  state $\ket{\psi_G^{(1)}(\phi,0,\pi/2)}$ (\ref{eq:2:13}) for different values of $\phi$, for qubit q[1]  with other qubits in  state $\ket{\psi_G^{(2)}(\phi,0,\pi/2)}$ (\ref{eq:2:14}) (c) and for qubit q[0]  with other qubits in  state   $\ket{\psi_G^{(3)}(\phi,0,\pi/2)}$ (\ref{eq:2:15}) (d) for different values of $\phi$.}
	\label{fig:8}
\end{figure}

We quantify  geometric measure of entanglement of qubit $q[0]$ corresponding to the vertex with degree 1 with other qubits in the  states $\ket{\psi_G^{(1)}(\pi,0,\theta)}$, $\ket{\psi_G^{(1)}(\phi,0,\pi/2)}$  for different values of $\theta$ (see Fig. \ref{fig:7} (a)) and  for different values of $\phi$ (see Fig. \ref{fig:8} (a)).  In addition, the entanglement of the qubit $q[1]$ corresponding to the vertex with degree 2 was quantified in the states $\ket{\psi_G^{(1)}(\pi,0,\theta)}$, $\ket{\psi_G^{(1)}(\phi,0,\pi/2)}$ (see Fig. \ref{fig:7} (b), Fig. \ref{fig:8} (b)). In order to detect the geometric measure of entanglement of qubits represented by vertices with  degrees 1 and 2 we choose qubits $q [0]$ and $q[1]$ because of small readout errors for these qubits in comparison with other qubits corresponding to  vertices with  the same degrees in the graph state $\ket{\psi_G^{(1)}(\phi,0,\theta)}$  (see Table \ref{trr1}).

 \begin{table}[!!h]
\caption{The calibration parameters of IBM's quantum computer  $\textrm{ibmq\_athens}$   on 9 June 2021 \cite{kk}.}\label{trr1}
\begin{center}
\begin{tabular}{ c c c c }
       & $Q_0$ &  $Q_1$ & $Q_2$ \\
Readout error ($10^{-2}$) & 1.07 & 1.30  & 1.70 \\
Gate error ($10^{-4}$)& 2.98 & 3.16 & 5.26 \\
       & $Q_3$ & $Q_4$ \\
Readout error ($10^{-2}$) & 1.31 & 2.00 & \\
Gate error ($10^{-4}$)& 2.54 & 2.89 &\\

CNOT error  ($10^{-3}$) & CX0$\_$1& CX1$\_$0 &  CX1$\_$2 \\
                            & 12.04 & 12.04& 11.13 \\
                            & CX2$\_$1 & CX2$\_$3 & CX3$\_$2\\
                            & 11.13 & 18.50 & 18.50 \\
                            & CX3$\_$4 & CX4$\_$3 \\
                            & 6.80 & 6.80 \\
\end{tabular}
\end{center}
\end{table}

 The geometric measure of entanglement of the qubit $q[1]$ corresponding to the vertex with degree 3 in the claw graph  with other qubits in the states $\ket{\psi_G^{(2)}(\pi,0,\theta)}$, $\ket{\psi_G^{(2)}(\phi,0,\pi/2)}$ was also calculated (see Fig. \ref{fig:7} (c), Fig. \ref{fig:8} (c)). To quantify the geometric measure of entanglement in the case of $n_l=4$  graph states $\ket{\psi_G^{(3)}(\pi,0,\theta)}$, $\ket{\psi_G^{(3)}(\phi,0,\pi/2)}$ represented by the complete graph were prepared using protocols (see Fig. \ref{fig:5} (c),  Fig. \ref{fig:6} (c)) and the geometric measure of entanglement of the qubit $q[0]$ with other qubits was quantified (see Fig. \ref{fig:7} (d), Fig. \ref{fig:8} (d)).

\vspace{5cm}

The results for the geometric measure of entanglement obtained on the quantum computer for qubits in the graph states $\ket{\psi_G^{(1)}(\pi,0,\theta)}$, $\ket{\psi_G^{(1)}(\phi,0,\pi/2)}$ (see  Fig. \ref{fig:7} (a), (b), Fig. \ref{fig:8} (a), (b)) are in good agreement with theoretical ones.  In the case of graph states $\ket{\psi_G^{(2)}(\pi,0,\theta)}$, $\ket{\psi_G^{(2)}(\phi,0,\pi/2)}$, $\ket{\psi_G^{(3)}(\pi,0,\theta)}$, $\ket{\psi_G^{(3)}(\phi,0,\pi/2)}$ the results for the entanglement are not in so good agreement with analytical results because to prepare
 these states  two-qubit gates have to be applied to qubits that are not connected directly according to connectivity map of IBM's quantum computer $\textrm{ibmq\_athens}$ (see Fig. \ref{fig:1}). Also, quantum protocols for preparing graph states represented by the complete graph (see Fig. \ref{fig:5} (c), Fig. \ref{fig:6} (c)) contain more gates than those for preparing graph states corresponding to the chain and the claw (see Fig. \ref{fig:5} (a), (b), Fig. \ref{fig:6} (a), (b)). This leads to accumulation of errors.

\section{Conclusion}

In the paper we have studied the geometric measure of entanglement of graph states generated by the action of controlled phase shift operators on a separable quantum state of the system, in which all the qubits are in arbitrary identical states (\ref{eq:2:1}). The expression for the geometric measure of entanglemet of a qubit with other qubits in  graph state (\ref{eq:2:1}) represented by an arbitrary graph has been found (\ref{eq:2:28}). We have concluded that the entanglement depends on the absolute values of the parameter of the phase shift operator $\phi$ and the parameter $\theta$ of states (\ref{eq:2:2}) as well as on the degree of the vertex representing the qubit in the graph.

 The geometric measure of entanglement has also  been calculated on the 5-qubit IBM's quantum computer $\textrm{ibmq\_athens}$. Graph states corresponding to the graph with the structure of quantum computer $\textrm{ibmq\_athens}$, the claw and the complete graphs have been prepared and the geometric measure of entanglement of qubits represented by vertices with degrees 1,2,3,4 has been found. Fixing parameter $\phi=\pi$ we have prepared graph states (\ref{eq:2:13})-(\ref{eq:2:15}) and quantified their entanglement for different values of $\theta$  (see Fig. \ref{fig:7}). In addition,  we have set $\theta=\pi/2$ and examined dependence of the geometric measure of entanglement for different values of the  parameter of the controlled phase gate $\phi$ in the case of graph states   (\ref{eq:2:13})-(\ref{eq:2:15}) (see Fig. \ref{fig:8}). The results obtained with quantum calculations on the  quantum device $\textrm{ibmq\_athens}$ are in good agreement with theoretical ones (see Figs. \ref{fig:7}, \ref{fig:8}).

\section{Acknowledgments}

The authors thank prof. Tkachuk V. M. for great support and useful comments during the research studies.
This work was supported by Project 2020.02/0196  (No. 0120U104801) from National Research Foundation of Ukraine.

\end{document}